\documentclass[useAMS,usenatbib]{mnras}
\usepackage{graphicx}
\usepackage{textcomp}
\usepackage{amssymb}\usepackage{hyperref}
\usepackage{amsmath,alltt}                                                                                                                     
\usepackage{multirow}
\usepackage{rotating}
\usepackage{lscape}
\usepackage{times}
\usepackage{hyperref}
\usepackage{pdflscape}
\usepackage{supertabular}

\newcommand\lsim{\mathrel{\rlap{\lower4pt\hbox{\hskip1pt$\sim$}}
        \raise1pt\hbox{$<$}}}
\newcommand\gsim{\mathrel{\rlap{\lower4pt\hbox{\hskip1pt$\sim$}}
        \raise1pt\hbox{$>$}}}
\newcommand{\be}{\begin{equation}}
\newcommand{\ba}{\begin{eqnarray}}
\newcommand{\ee}{\end{equation}}
\newcommand{\ea}{\end{eqnarray}}

\title[Constraints on BS formation in Galactic GCs]{Constraints on blue straggler formation mechanisms in Galactic globular clusters from proper motion velocity distributions}

\author[N. W. C. Leigh, T. Panurach, M. Simunovic, A. M. Geller, et al.]{N. W. C. Leigh$^{1,2,3}$, T. Panurach$^{1,4}$, M. Simunovic$^{5}$, A. M. Geller$^{6,7}$, D. Zurek$^{1}$, 
\newauthor
M. M. Shara$^{1}$, A. Sills$^{8}$, C. Knigge$^{9}$, N. M. Gosnell$^{10}$, R. Mathieu$^{11}$, T. H. Puzia$^{12}$, 
\newauthor
J. Ventura$^{1,4}$, Q. Minor$^{1,13}$ \\
$^{1}$Department of Astrophysics, American Museum of Natural History, New York, NY 10024, USA\\
$^{2}$Department of Physics and Astronomy, Stony Brook University, Stony Brook, NY 11794-3800, USA\\
$^{3}$Center for Computational Astrophysics, Flatiron Institute, 162 Fifth Avenue, New York, NY 10010, USA\\
$^{4}$Department of Physics and Astronomy, Hunter College, City University of New York, 695 Park Avenue, New York, NY 10065, USA \\
$^{5}$Gemini Observatory, 670 N. A’ohoku Pl, Hilo, HI 96720, USA \\
$^{6}$Center for Interdisciplinary Exploration and Research in Astrophysics (CIERA) and Department of Physics and Astronomy, \\ Northwestern University, 2145 Sheridan Rd, Evanston, IL 60208, USA \\
$^{7}$Adler Planetarium, Dept.\ of Astronomy, 1300 S. Lake Shore Drive, Chicago, IL 60605, USA \\
$^{8}$McMaster University, Department of Physics and Astronomy, Hamilton, Ontario, Canada, L8S 4M1\\
$^{8}$Department of Physics, Queensborough Community College, City University of New York, Bayside, NY 11364, USA \\
$^{9}$University of Southampton, School of Physics and Astronomy, Southampton SO17 1BJ, UK\\
$^{10}$Colorado College, Department of Physics, 14 E. Cache La Poudre St, Colorado Springs, CO 80903, USA\\
$^{11}$Department of Astronomy, University of Wisconsin - Madison, 475 North Charter Street, Madison, Wisconsin 53706, USA\\
$^{12}$Institute of Astrophysics, Pontificia Universidad Cato ́lica de Chile, Av. Vicun\ ̃a Mackenna 4860, 7820436 MaculSantiago, Chile \\
$^{13}$Department of Science, Borough of Manhattan Community College, City University of New York, New York, NY 10007, USA}

\begin{document}

\date{Accepted. Received; in original form}

\pagerange{\pageref{firstpage}--\pageref{lastpage}} \pubyear{2008}

\maketitle

\label{firstpage}

\begin{abstract}
For a sample of 38 Galactic globular clusters (GCs), we confront the observed distributions of blue straggler (BS) proper motions and masses (derived from isochrone fitting) from the BS catalog of Simunovic \& Puzia with theoretical predictions for each of the two main competing BS formation mechanisms.  These are mass transfer from an evolved donor on to a main-sequence (MS) star in a close binary system, and direct collisions involving MS stars during binary encounters.  We use the \texttt{FEWBODY} code to perform simulations of single-binary and binary-binary interactions.  This provides collisional velocity and mass distributions for comparison to the observed distributions.  Most clusters are consistent with BSs derived from a dynamically relaxed population, supportive of the binary mass-transfer scenario. In a few clusters, including all the post-core collapse clusters in our sample, the collisional velocities provide the best fit.
%We further calculate spectral energy distributions (SEDs) for our BS samples in all 38 GCs, for comparison to synthetic single star SEDs.  We use the \texttt{synphot} package to calculate fluxes in all five photometric bands.  We search for excesses in our bluest and reddest filters due to, respectively, hot young white dwarf binary companions and MS companions.  These are smoking guns of, respectively, recent mass transfer and binary-mediated collisions.  We identify those GCs for which the fraction of BSs consistent with having a flux excess exceeds the corresponding fractions for control samples of red giant branch and MS turn-off stars.
%The results of this analysis place direct constraints on the fraction of BSs that are actually unresolved binary star systems.  Our results suggest a BS binary fraction $> 50\%$ in every cluster in our sample, which is significantly higher than inferred photometrically for MS-MS binaries.  However, comparisons to a control sample of main-sequence turn-off stars suggests that caution should be applied when interpreting these results, since the detection of any binary companion is sensitive to both the absolute magnitude of the target object and the corresponding uncertainty.
%This represents a direct test of the collisional hypothesis for BS formation, which predicts that the BS-MS binary fraction should be high.  
%We also constrain the preferred mass ratios of any BS binaries, and generate mass ratio distributions for detected BS binaries in all GCs in our sample.  
%We find a mean BS-MS binary fraction of xx\% ...
\end{abstract}


%%%%%%%%%%%%%%%%%%%%%%%%%%%%%%%
\section{Introduction}

In Galactic globular clusters (GCs), the central densities are so high ($>$ 10$^5$ M$_{\odot}$ pc$^{-3}$) that direct encounters between single and binary stars occur frequently.  The exact rate depends on the host cluster properties but, for the densest GCs, the time-scale for single-binary and binary-binary (note that these two timescales are roughly equal for binary fractions $\sim$ 10\% \citep{sigurdsson93,leigh11}) encounters to occur is of order 1-10 Myr \citep[e.g.][]{leigh11,leigh13,geller15}.  Furthermore, simulations have shown that a large fraction of these interactions should produce direct collisions between two or more main-sequence stars \citep[e.g.][]{leonard89,leigh12,leigh13,hypki13}.  For interaction parameters typical of GCs, single-binary and binary-binary interactions should produce direct MS-MS collisions of order a few to a few tens of percent of the time \citep{leigh12}.  The corresponding main-sequence lifetimes of these collision products are poorly known from stellar evolution theory, but should be of order $\sim$ 1 Gyr \citep{sills97,sills01}.  This implies that approximately (1 Gyr)/(1-10 Myr) $\times$ 10\% $\sim$ 10-100 direct collisions should be present in a dense GC at any time.

\begin{figure}
\begin{center}
\includegraphics[width=\columnwidth]{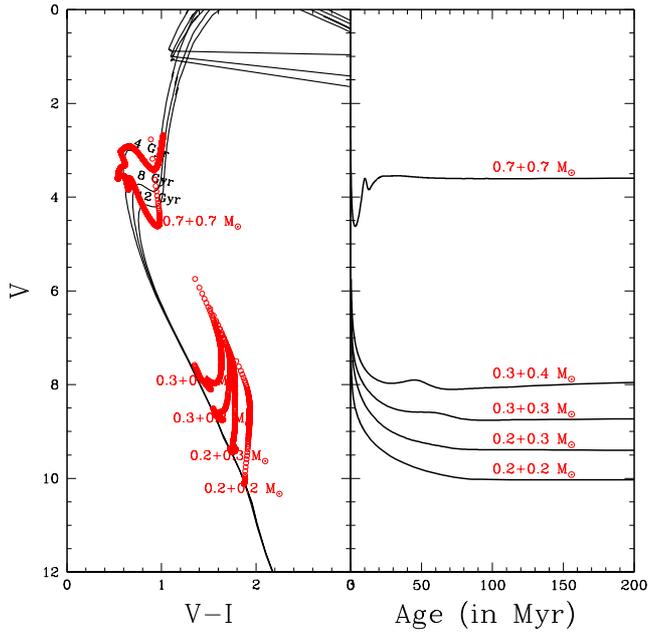}
\end{center}
\caption[Stellar evolution tracks for collision products of various masses]{Stellar evolution tracks calculated with the YREC code \citep{sills97,sills01} are shown in the $(V-I)-V$-plane for the collision products of main-sequence stars of the indicated mass (red; left panel).  The red points are plotted every 0.1 Myr for the first $\sim$ 10 Myr before the time-step increases progressively from this value to $\sim$ 100 Myr at an age of about $\lesssim$ 0.1 Gyr.  The solid black lines show theoretical isochrones from \citet{dotter10} at cluster ages of 4, 8 and 12 Gyr.  The time evolution of the V-band magnitude of each collision product is also shown in the right panel in Myr; collisions begin much brighter and redder than the zero-age main-sequence (ZAMS; left panel), but quickly dim and contract back down to become indistinguishable from MS stars in a cluster CMD (right panel).     
\label{fig:fig1}}
\end{figure}

So where are all these collision products?  Observationally identifying collision products is anything but straight forward.  Arguably the most promising candidates are blue straggler (BS) stars, which are sources that appear brighter and bluer than the main-sequence turn-off (MSTO) in a cluster colour-magnitude diagram (CMD; see Figure~\ref{fig:fig1}).  Two primary channels for BS formation have been proposed; (previous and/or ongoing) mass transfer from an evolved donor on to a main-sequence star in a close binary star system \citep[e.g.][]{mccrea64,knigge09,leigh11,gosnell14,gosnell15}, and direct collisions involving main-sequence stars that are typically mediated via resonant interactions involving binaries \citep[e.g.][]{hills75,shara97,leigh13,hypki13}.  Other possible, albeit related, formation mechanisms include mergers of close or compact MS-MS binaries, and Lidov-Kozai-induced mergers of the inner MS-MS binaries of hierarchical triple star systems \citep[e.g.][]{perets09}.  We include both of these formation mechanisms in the ``collisional'' channel for BS formation, since observationally they should be indistinguishable from BSs formed via MS-MS collisions during single-binary and/or binary-binary interactions.

Lower-mass collision products that end up further down the main-sequence could also be identifiable photometrically in a cluster CMD.  As shown in the left panel of Figure~\ref{fig:fig1}, immediately after the collision, the product is out of thermal equilibrium, inflated and has evolved up the Hayashi track in the CMD \citep[e.g.][]{sills97,sills99,sills01}.  For at least a brief period of time, this places the collision products towards red colours and brighter luminosities than the MS, where it could potentially be less ambiguous to identify.  However, this phase in the lifetime of the collision product is brief (see Figure~\ref{fig:fig1}), and distinguishing these objects from unresolved binary and/or triple star systems in clusters may be difficult.

Observationally, the best evidence for collisions in GCs comes from direct measurements of individual blue stragglers.  For example, \citet{shara97} reported the discovery of a BS with a mass of $\sim$ 1.7 $\pm$ 0.4 M$_{\odot}$ in the Galactic GC 47 Tuc, consistent with being nearly twice the mass of the MSTO.  The authors argued that such a massive BS could only have formed if at least one collision or merger occurred.  Less encouragingly, \citet{leigh07}, \citet{knigge09} and \citet{leigh13} argued that the dominant BS formation mechanism operating in GCs is binary mass transfer.  
%These authors found little to no correlation between the properties of a large sample of BSs in $\sim$ 60 GCs and the cluster collision rate, but a strong correlation with the core mass, as expected if most BSs are descended from binary stars.  
 
Theoretically, simulations have shown that \textit{both} collisions and binary mass transfer can produce BSs in significant numbers \citep[e.g.][]{hurley05,geller13b,chatterjee13,hypki13}.  Interestingly, \citet{chatterjee13} found from a suite of Monte Carlo simulations for GC evolution that BSs formed from collisions tend to be more massive than BSs formed from binary mass transfer.  Consequently, collisional BSs have shorter main-sequence lifetimes than mass transfer BSs in their simulations.  This contributes to a decrease in the probability of actually observing collision products in the BS region of the cluster CMD, relative to mass transfer BSs.  Moreover, it is not always straight-forward to cleanly assign one formation mechanism to all BSs \citep{hypki13}.  For example, \citet{hurley05} and \citet{geller13b} found from $N$-body simulations of star cluster evolution that many BSs have very complicated formation histories, experiencing multiple dynamical exchange interactions, episodes of mass transfer and even collisions.  
 
Several studies have looked for evidence of a collisional origin for BSs in GCs \citep[e.g.][]{sills97,sills01}.  Most of these previous studies used indirect methods, such as looking for correlations between BS population size and collision rate \citep{piotto04,leigh07}.  Based on such an approach, \citet{knigge09} argued that most BSs have a binary origin, since their abundance is not strongly correlated with the cluster collision rate in the cores of Galactic GCs, but is strongly correlated with the core masses.  Moreover, most BSs in old open clusters seem to have a mass transfer origin \citep{mathieu09,geller11,gosnell14,gosnell15}, as evidenced by the detection of hot white dwarf and post-AGB/HB companions to these BSs.  

This highlights an important commonality of the two competing formation mechanisms for BS production:  they each predict that BSs should most likely have a binary companion.  The nature of the companion depends on the formation mechanism; mass transfer predicts white dwarf companions \citep[e.g.][]{geller11}, whereas MS-MS collisions during single-binary and/or binary-binary interactions predict MS or old white dwarf companions \citep[e.g.][]{leonard89,leigh11}.\footnote{We note that the probability of obtaining a white dwarf companion is also significant within the framework of the collisional hypothesis for BS formation, since GCs are comprised of old stars with a non-negligible mass-to-light ratio of order two.  Hence, this mechanism predicts \textit{mostly} MS binary companions to BSs, but with some old WDs as well.  These WDs should be sufficiently old and have cooled enough that they should be difficult if not impossible to detect.  It follows that the BS companions to these old WDs should appear as isolated single BSs photometrically.}  

If theoretical simulations predict the formation of BSs via dynamical collisions/mergers, where are all the observed main-sequence collision products in Galactic GCs?  Here, we address this question directly using multi-epoch multi-band photometry from the Hubble Space Telescope (HST).  This is done using the blue straggler catalog of \citet{simunovic16}, which provides proper motion-cleaned BS samples for 38 Galactic GCs.  
%We also include in our candidate collision samples any low-mass collision products we identify, defined as any likely cluster members (based on the their proper motions and an analysis of the HST images directly) with a photometric appearance that puts it above and to the red of the equal-mass triple sequence in the (V-I)-I CMD (see Figure~\ref{fig:fig1}).  Low-mass collision products are the only cluster members that can populate this region of the CMD.  
We compare the observed BS velocity (from our 2-D proper motions) and mass distributions to theoretical predictions for each of the two main competing formation scenarios for BS formation.  

\begin{figure}
\begin{center}
\includegraphics[width=\columnwidth]{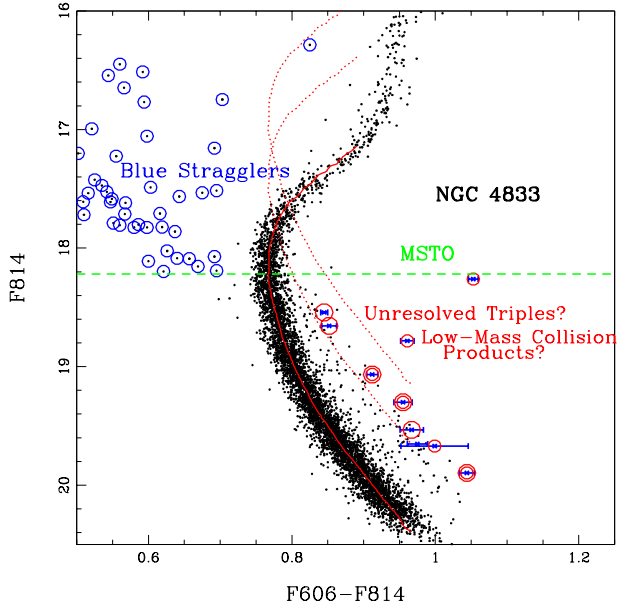}
\end{center}
\caption[Example CMD for the GC NGC 4833]{Example colour-magnitude diagram for the GC NGC 4833, delineating the BSs and low-mass collision product candidates.  The red circles indicate candidate low-mass collision products that do not have any nearby bright giants in the images, and are hence secure from contamination from giants.  The green dashed horizontal line indicates the I-band magnitude corresponding to the main-sequence turn-off, found from isochrone-fitting \citep{dotter10}.  The solid red line shows the fiducial MS.  The dashed red lines show the fiducial MS shifted vertically by 0.75 and 1.25 mag; these lines correspond to, respectively, the equal-mass binary and triple sequences.  We note that the photometric errors in the I-band are smaller than the sizes of the plotted data points for each low-mass collision product candidate.  
\label{fig:fig2}}
\end{figure}

Additionally, this study is intended as a first step toward addressing the over-arching question:  Where are all the collision products in Milky Way globular clusters?  We will consider low-mass collision products appearing below the main-sequence turn-off in the cluster CMD (see Figures~\ref{fig:fig1} and~\ref{fig:fig2}) in a future paper.  These low-mass collision products should spend some of their lives sufficiently far off the main-sequence that they might potentially be observable above the equal-mass binary line, where very few if any multiple star systems should exist (with the possible exception of unresolved triple star systems composed of three nearly identical main-sequence stars).

%In Section~\ref{model}, we begin by calculating the orbital separation beyond which an interloping single star will likely disrupt a BHB.  Next we calculate the rate of encounters between BHBs and single objects in both the disk and spherical components. Then we derive an analytic expression for the mean number of scattering interactions $N_{\rm GW}$ required to harden a BHB until the mean encounter timescale exceeds the inspiral timescale from GW emission. Finally we compare our analytic estimate for $N_{\rm GW}$ to the results of Monte-Carlo simulations that account for inter-encounter hardening and circularization due to GW emission.  Section~\ref{discussion} discusses and summarizes our results.

\section{Analysis} \label{method}

In this section, we present the multi-epoch, multi-band proper motion-cleaned photometry used to obtain the catalogs of blue stragglers, and hence collision product candidates, used in this study. We further present the numerical scattering experiments used to generate theoretical predictions for the velocity and mass distributions of collision products formed during single-binary and binary-binary interactions.
%, as well as the synthetic photometry used for comparison to the observed spectral energy distributions. 

\subsection{The data} \label{data}

The Ultraviolet-Visible Channel (UVIS) camera on board the Hubble Space Telescope (HST) surveyed almost half (out of over 150) of the Galactic globular cluster population in near-UV passbands \citep{piotto15}. These data, together with an analogous optical survey performed previously using the Advanced Camera for Surveys (ACS) camera on board HST \citep{sarajedini07}, provides the input for multi-epoch photometry. Cluster members have been selected in \citet{simunovic16} using HST proper-motions, removing in the process the vast majority of foreground and background contaminants from the field of view. This now allows for a much more secure photometric identification of blue stragglers and multiple star systems in the proper-motion cleaned CMDs. 

%\subsubsection{Blue straggler stars} \label{BSs}
%
%We obtain velocities for our BSs from the proper motion-cleaned sample of \citet{simunovic16}.  These represent 2-D velocities, projected in the plane of the sky.  BS masses are obtained from isochrone-fitting to the observed CMDs.  We assume 1 Gyr isochrones \citep{dotter10} and adopt cluster metallicities from \citet{harris96}.  For this experiment, we assume that all BSs are isolated single stars (or, equivalently, binary systems with a less luminous companion that does not contribute to the observed photometry), but later relax this assumption to constrain the possible presence of a binary companion.  Consequently, we emphasize that, if our observed BSs are indeed members of binary star systems with companions that affect the observed photometric appearance of the BSs, these derived masses will be incorrect.  We will return to this in Section~\ref{discussion}.

\subsubsection{BS proper-motion velocities and mass estimates} \label{BSs}

The stellar proper motions represent 2-D motions, projected in the plane of the sky. We transform the relative proper motions into velocity units by using the GCs' heliocentric distances from \cite{harris96} to convert to a physical distance, and then dividing by the time baseline of the multi-epoch data.  Therefore, this way we effectively obtain the relative velocity of each BS with respect to the GC mean motion.  As discussed already in \citet{simunovic16}, the derived velocities can be subject to large uncertainties (e.g. for the GCs more distant to Earth, in which cases the true stellar motions are buried in the instrumental astrometric errors). For this reason, the BSs used for the comparison with the scattering experiments are selected using the condition that their calculated proper motion error must be smaller than a characteristic maximum angle. The latter is defined as the astrometric projected 2-D displacement, at the distance of each cluster, calculated for the time equal to the multi-epoch baseline and the central velocity dispersion of each parent cluster. The central velocity dispersion for each GC is calculated from a dynamical mass approach, where the total cluster mass is computed from the V-band mass-to-light ratios from the SSP models of \cite{bruzual03}, for each GC metallicity assuming a 12 Gyr old stellar population. The V-band magnitudes and metallicities are taken from \cite{harris96}. The aforementioned condition reduces our sample size but ensures that the measured BS velocities used in the analysis are not dominated by the astrometric errors.  Moreover, for the resulting BS samples we do not find any evidence of a bias in luminosity nor in cluster-centric distance, as found via KS tests.

Motivated by the opportunity to probe the effect of stellar mass in our scattering experiments, we use a simple approach to obtain stellar mass estimates for our BS sample. The method uses the position of the BS in the CMD with respect to a set of appropriate isochrones, and finds the interpolated mass value that can reproduce the location of each BS in the CMD. We use a set of Dartmouth isochrones \citep{dotter10} and adopt the corresponding cluster parameters of reddening $E(F606W-F814W)$, metallicity and distance modulus from the values reported in \cite{dotter10}, or instead from \cite{harris96} for the GC values not available from the former paper. We convert the galactic extinction factors given in \cite{harris96} to the ACS/WFC filter system by using the transformations from \cite{sirianni05}. For all GCs, we adopt a representative value of 0.2 for the [$\alpha$/Fe] enhancement. 
%The BS selection method for our catalogs described in \cite{simunovic16} used a 1 Gyr old isochrone\footnote{We note that the obtained BS masses are not very sensitive to this choice of isochrone age for the blue limit.} as the blue limit for the BS region, therefore 

We use a set of isochrones with a range in ages from 1 to 9 Gyr as the stellar library grid that fully encompasses the BS region in the CMD. This approach has been used in previous studies \citep{ferraro06,lanzoni07,lovisi12} in an attempt to characterize the BS physical properties in individual GCs, but it should be noted that this approach assumes all BSs are isolated stars or, equivalently, binary systems with a less luminous companion that does not contribute to the observed photometry. Consequently, we emphasize that, if our observed BSs are indeed members of binary star systems with companions that affect the observed photometry, these derived masses will be representative of the integrated light, but will be incorrect for individual components. %We will return to this in Section~\ref{synth}, where we perform a more detailed analysis involving SED-fitting to constrain the binary nature of our BSs.

Importantly, we do not infer much about the observed BS sample from any analysis of these mass distributions, and instead only compare between our theoretical predictions for collisions during 1+2 and 2+2 interactions.  This is because it suffers from not knowing the binarity of each BS in our samples.  Regardless, the inferred mass can be used as a proxy for the total luminosity, and hence should correlate strongly with the underlying binarity of a given sample.   %\textbf{If, as described above, a more detailed and thorough SED-analysis is performed that can reveal the single or binary nature of each BS in our sample, this would allow us to return to this component of our analysis and re-perform it with significant refinements informed by our initial analysis.  We hope to perform this SED-fitting exercise to constrain the binarity of each BS in a forthcoming paper.}

%One of the physical effects that can degrade the photometry in the CMD (see Figure~\ref{fig:fig2}) is differential reddening (DR) across the field of view.  Given that BSs are much brighter and bluer than are low-mass collision products, the effects of differential reddening should be much less significant, and can be neglected.  However, for low-mass collision products appearing above and to the red of the MS locus in the cluster CMD (but below the subgiant branch), the photometry is more sensitive to the effects of DR, such that it should be corrected for when we return to this sample in a forthcoming paper.

\subsection{Numerical scattering simulations} \label{simulations}

In this section we present results from roughly a million numerical scattering experiments of single-binary (1+2) and binary-binary (2+2) interactions performed with the \texttt{FEWBODY} code \citep{fregeau04}. We choose parameters for these scattering experiments that are relevant for the core of each GC in our sample, using the observed total cluster masses and half-mass radii \citep{harris96}.  In order to sample the masses of the stars for input to \texttt{FEWBODY}, we sample the masses of the stars from the observationally-constrained present-day stellar mass function for that cluster.  This is done using the corresponding relations provided in \citet{leigh12}, adopting the total cluster mass from \citet{harris96}, calculated assuming $M/L = 2$, appropriate for an old stellar population, and the MSTO mass obtained from fitting theoretical isochrones.  Our isochrones are taken from \citet{dotter10}, adopting the cluster metallicities and extinctions provided in \citet{harris96} and a cluster age of 12 Gyr.

For a given binary, we first choose the primary mass ($m_{\rm 1}$) from the same mass function. We then draw a mass ratio ($q = m_{\rm 2}/m_{\rm 1}$) from a uniform distribution to select the secondary mass ($m_{\rm 2}$), enforcing the criteria that $q \le 1$ and $m_{\rm 2} > 0.1 M_{\odot}$.  We derive a radius for each star using this stellar mass and the cluster [Fe/H], following the method of \citet{tout96}.  We provide these stellar parameters to \texttt{FEWBODY} such that it can identify physical collisions during the encounters, using the ``sticky-star'' approximation.  This assumes that a collision occurs if the stellar radii overlap at any point during the interaction (see \cite{fregeau04} and \cite{leigh12} for more details).  We choose orbital elements from the observed distributions of binaries with solar-type primary stars in the Galactic field from \citet{raghavan10}, which are also consistent with observations of solar-type binaries in open clusters \citep[e.g.][]{geller13a,geller12}.  Hence, we draw orbital periods from a log-normal distribution with a mean of $\log(P {\rm [days]}) =$ 5.03 and $\sigma =$ 2.28.  The short period limit is equivalent to the one at the Roche radius \citep{eggleton83} and the long-period limit at the hard-soft boundary of the core. This translates to detached hard binaries initially, and soft binaries that are disrupted promptly.  We estimate the maximum period for a given hard binary, $P_{\rm hs}$, using the virial theorem.  This means that the velocity at infinity for the incoming object is drawn from a lowered Maxwellian distribution (typical for star clusters), which is defined by the velocity dispersion and the escape velocity, both calculated at the center of a Plummer model (with the observed cluster mass and half-mass radius).  Finally, we draw orbital eccentricities from a uniform distribution.

All angles that uniquely define each encounter are drawn randomly from an isotropic distribution.  We choose the impact parameter for a given encounter randomly from a uniform distribution between 0 and 1 times the binary semi-major axis in a 1+2 encounter, or the sum of the two binary semi-major axes for a 2+2 encounter.  The above parameters define discrete 1+2 and 2+2 scattering experiments that are appropriate for the cores of typical Milky Way star clusters.  

For each GC in our sample, we perform 10$^4$ 1+2 and the same number of 2+2 unique numerical scattering experiments.  We project the resulting 3-D velocity distributions for our collision products on to the plane of the sky, in order to obtain 2-D velocity distributions for direct comparison to our observed 2-D BS velocities.

\section{Results} \label{results}

In this section, we present and discuss the results of our study to help constrain the formation mechanism of BSs in Galactic GCs. This is accomplished by comparing the observed BS velocity and mass distributions to the predictions of theory.
%In this section, we present and discuss the results of our photometric search for the nature of any binary companions to blue stragglers in the CMDs of 38 Galactic GCs.  We begin by comparing the observed BS velocity and mass distributions to the predictions of theory, before moving on to comparing the observed BS spectral energy distributions to theoretical templates.

\subsection{Blue straggler velocities and masses} \label{mandv}

\begin{figure}
\begin{center}
\includegraphics[width=\columnwidth]{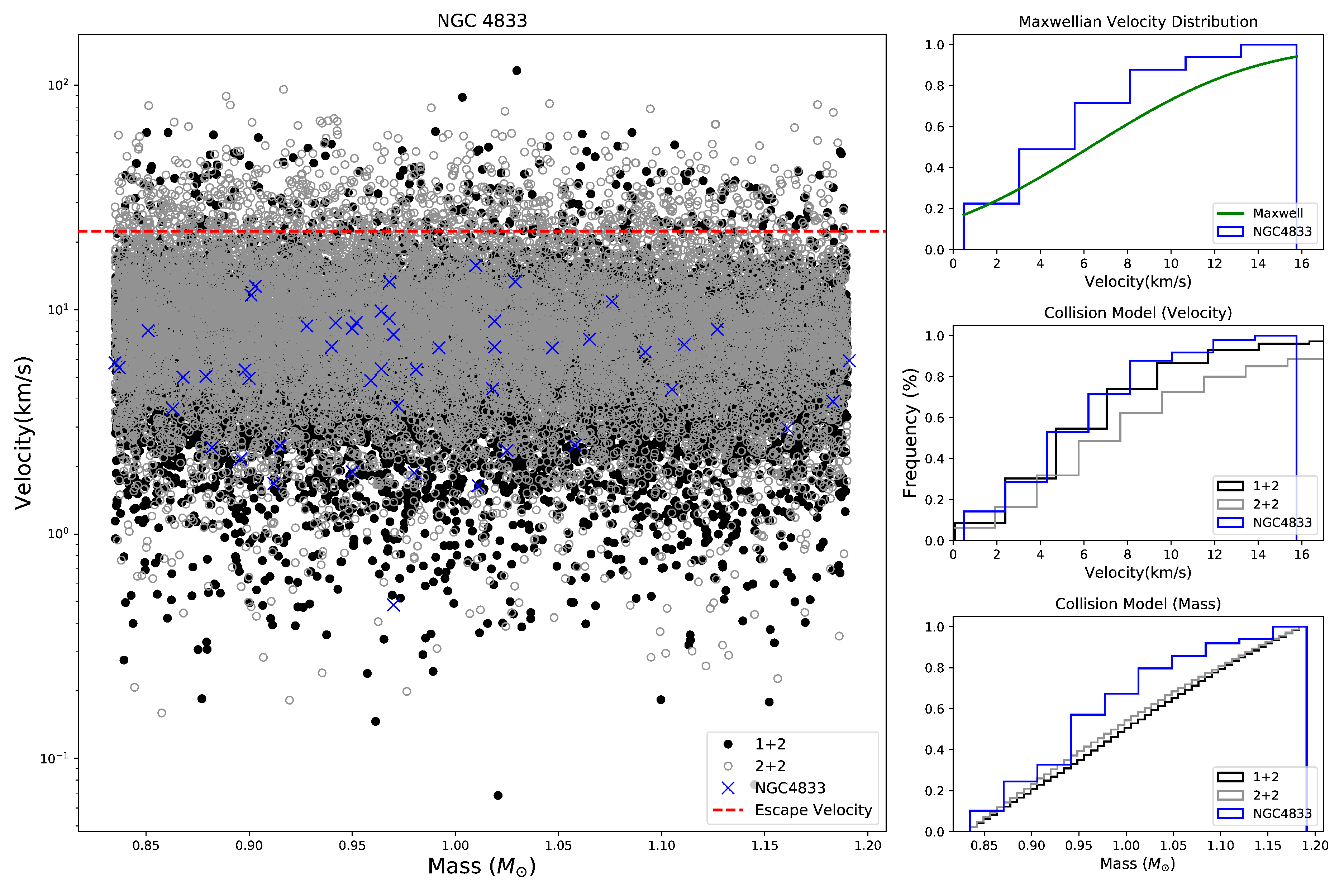}
\end{center}
\caption[Comparison between the observed and simulated 2-D BS velocity and mass distributions for NGC4833]{Comparison between the observed and simulated 2-D BS velocity and mass distributions for NGC4833.  (\textit{Left panel}): The velocities are plotted on the y-axis, and the collision product masses on the x-axis.  Blue points correspond to the observed BS data, whereas the solid black and light open grey circles correspond to 1+2 and 2+2 interactions, respectively.  The dashed red line delineates the cluster escape velocity from the core.  (\textit{Right panels}): We show histograms of the observed BS mass (bottom panel) and velocity distributions (top two panels).  In the bottom panel, the derived BS masses are compared directly to theoretical predictions for the collisional hypothesis during both 1+2 and 2+2 interactions obtained using our numerical simulations.  In the middle panel, the same exercise is performed, but with the observed and simulated velocity distributions.  In the top panel, the observed BS velocity distributions are compared to a Maxwellian velocity distribution, calculated assuming an average object mass equal to twice the mass of an average MS single star (i.e., $\sim$ 0.5 M$_{\odot}$).  We note that many of the black points are not seen, since the open grey circles have been over-plotted.  This is because the black points occupy a smaller region in this parameter space relative to the grey points.
\label{fig:fig7}}
\end{figure}

\begin{figure}
\begin{center}
\includegraphics[width=\columnwidth]{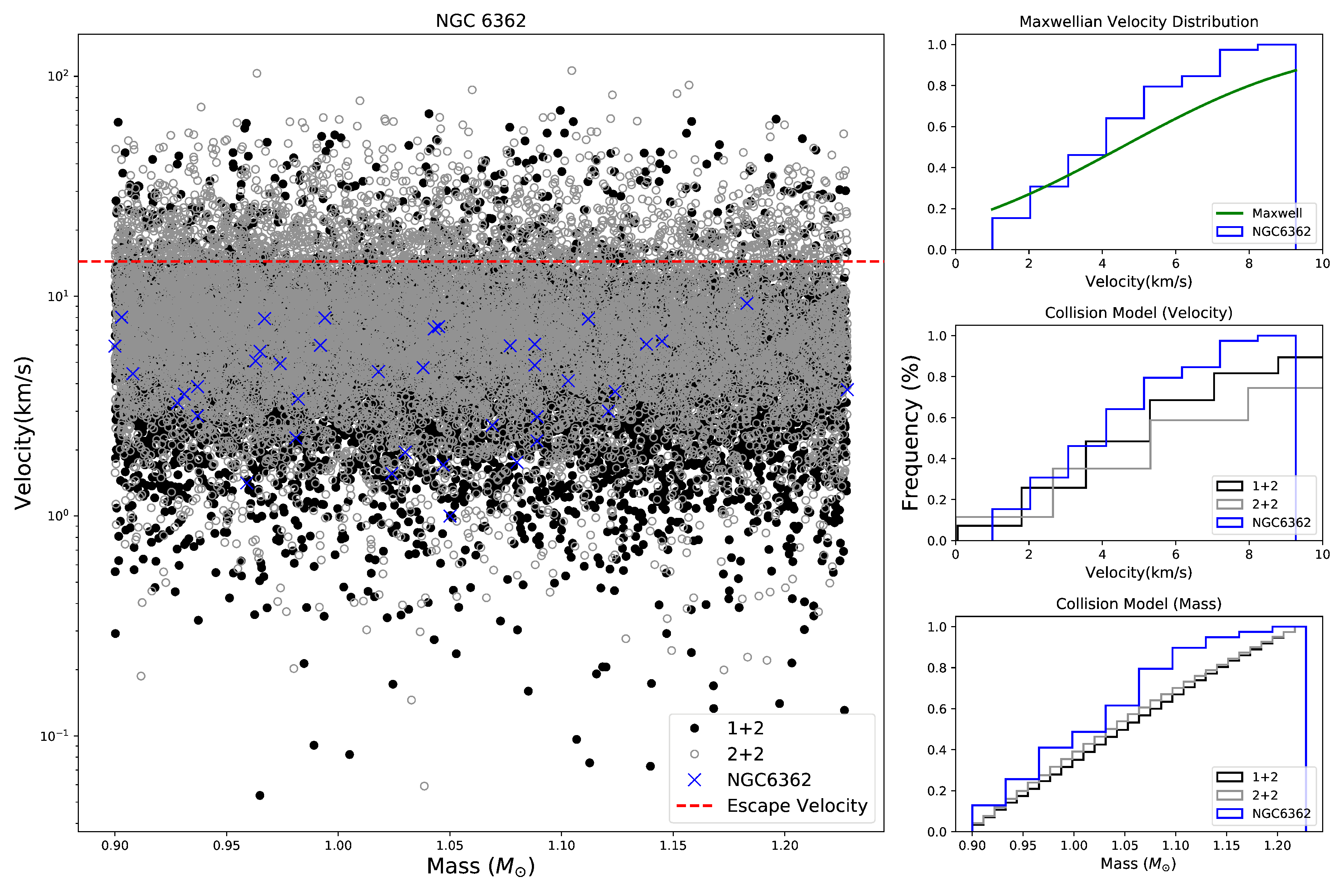}
\end{center}
\caption[Comparison between the observed and simulated 2-D BS velocity and mass distributions for NGC6362]{Same as Figure~\ref{fig:fig7}, but for the observed BS population in the GC NGC6362.
\label{fig:fig8}}
\end{figure}

\begin{figure}
\begin{center}
\includegraphics[width=\columnwidth]{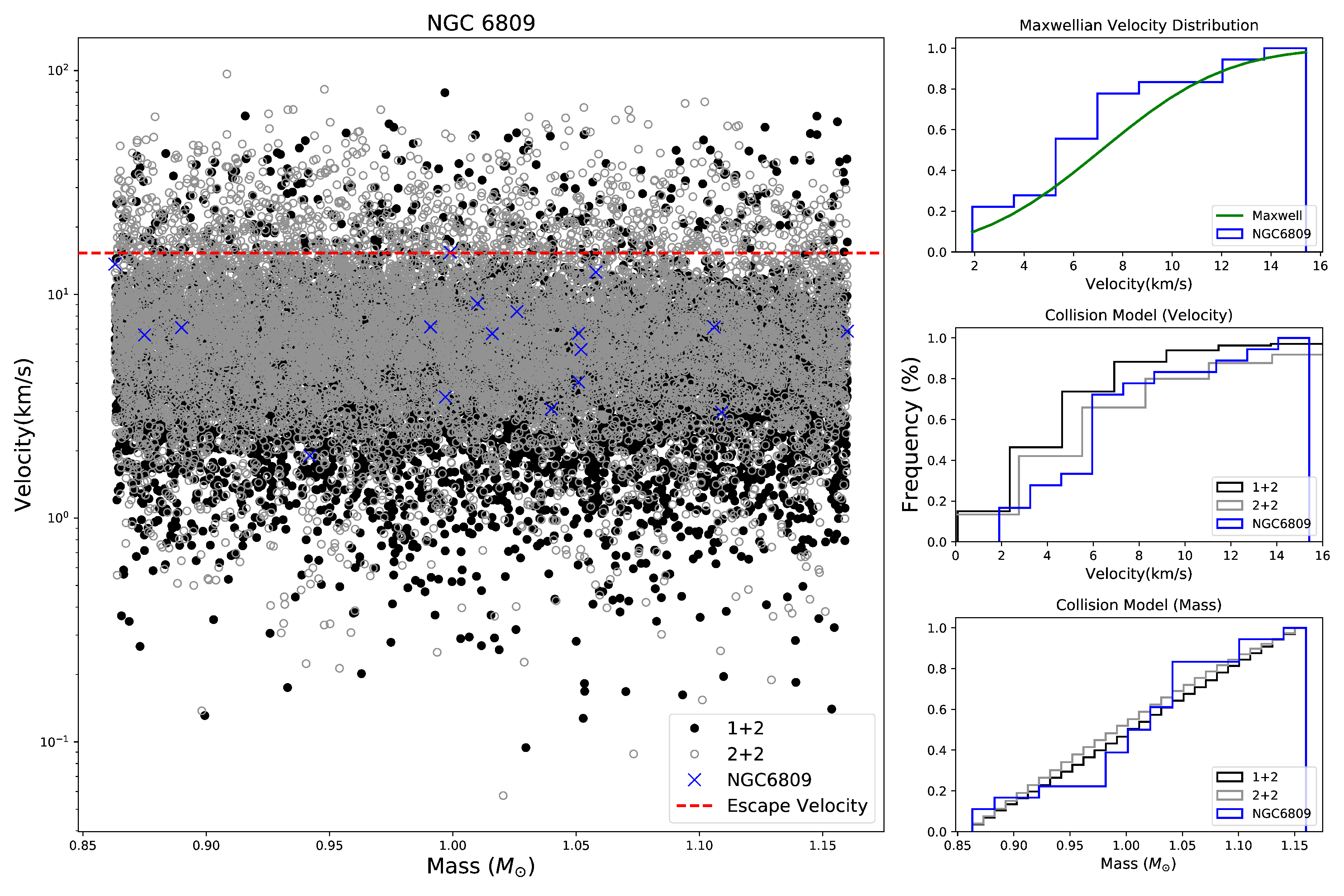}
\end{center}
\caption[Comparison between the observed and simulated 2-D BS velocity and mass distributions for NGC6809]{Same as Figure~\ref{fig:fig7}, but for the observed BS population in the GC NGC6809.
\label{fig:fig9}}
\end{figure}

\begin{figure}
\begin{center}
\includegraphics[width=\columnwidth]{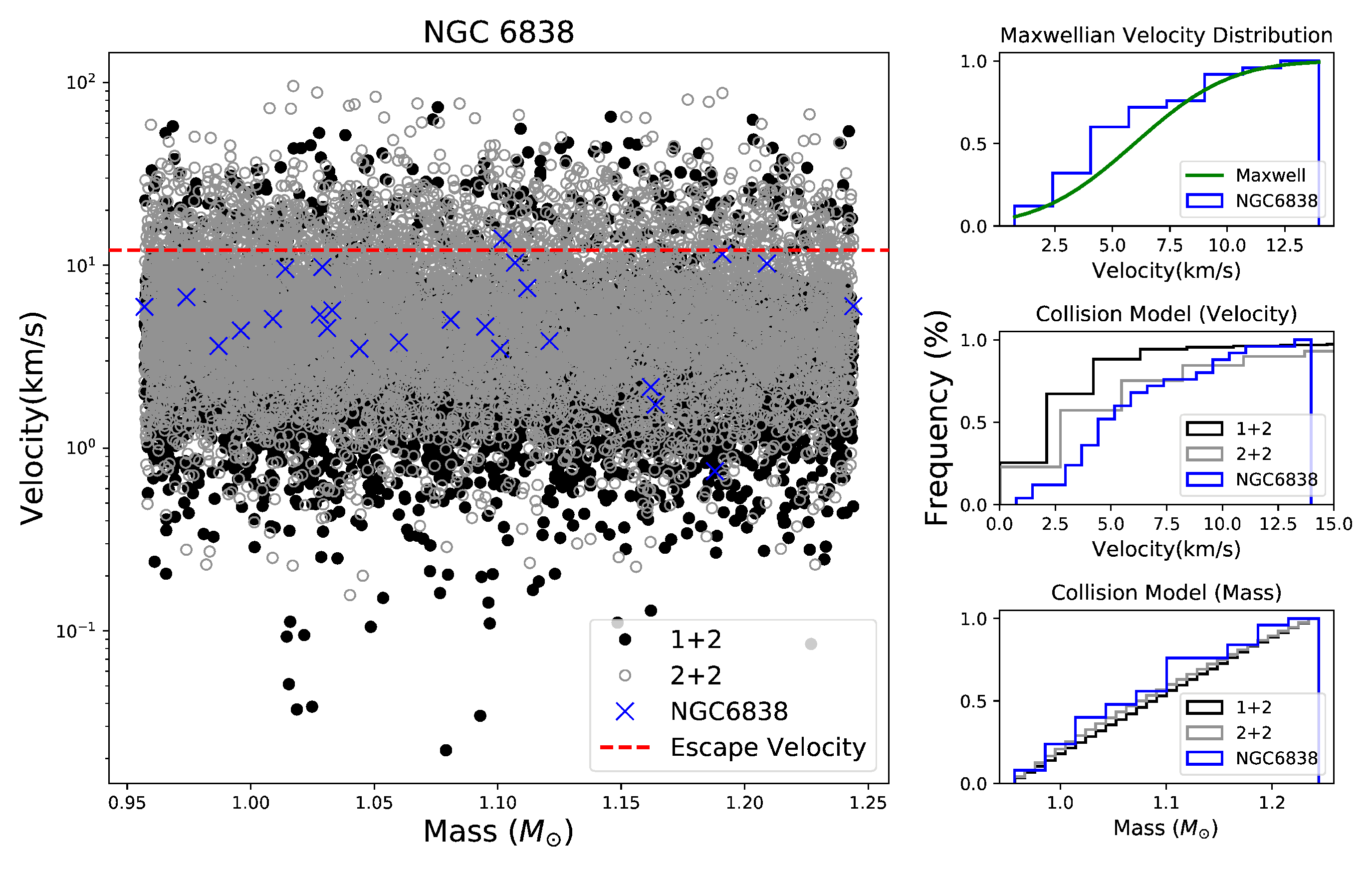}
\end{center}
\caption[Comparison between the observed and simulated 2-D BS velocity and mass distributions for NGC6838]{Same as Figure~\ref{fig:fig7}, but for the observed BS population in the GC NGC6838.
\label{fig:fig10}}
\end{figure}

\begin{figure}
\begin{center}
\includegraphics[width=\columnwidth]{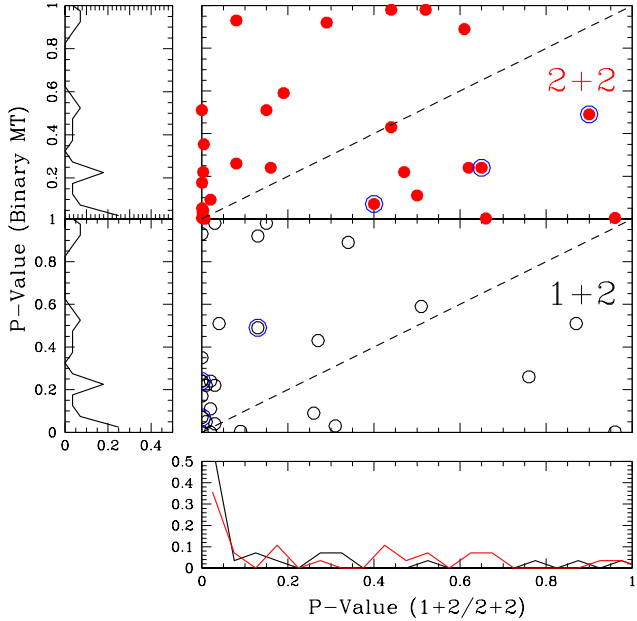}
\end{center}
\caption[Comparison between the observed and simulated 2-D BS velocity distributions for all GCs]{The $p$-values from our KS tests are plotted for both the binary mass transfer hypothesis (y-axis) and the 1+2/2+2 collisions hypotheses (x-axis; bottom/top inset, respectively).  The open black and filled red circles correspond to, respectively, 1+2 and 2+2 collisions.  The open blue circles denote the post-core collapse clusters in our sample.  The straight dashed line shows the dividing line between the competing formation scenarios; clusters whose data points fall in the upper left quadrant prefer binary mass transfer, whereas clusters that fall in the lower right quadrant prefer 1+2/2+2 collisions.  
\label{fig:fig11}}
\end{figure}

Here we present the results of comparing the observed BS velocities and masses, as derived in Section~\ref{BSs}, to the theoretical predictions for the competing BS formation mechanisms, namely i) MS-MS collisions during single-binary (1+2) or binary-binary (2+2) interactions and ii) binary mass transfer.  The latter are calculated using a suite of numerical scattering experiments performed with the \texttt{FEWBODY} code (see \cite{fregeau04} for more details), as described in Section~\ref{simulations}.  For the former, we assume a Maxwellian velocity distribution for an average or typical binary population in the cluster, i.e., assuming that all binaries have masses equal to twice the mean stellar mass.  We are only able to compare this theoretical expectation to the observed velocity data but not the mass distribution, since at least for the mass-transfer and/or binary-merger hypothesis a clear theoretical prediction is lacking.  Moreover, we note that only 29 GCs have BS velocity samples. This is because we chose not to include GCs whose BS velocity sample, after applying the selection criteria described in Section~\ref{BSs}, included less than 8 stars, to ensure statistical significance. The main results of our analysis are summarized below in Table~\ref{table:stats} and Figures~\ref{fig:fig7}-\ref{fig:fig11}.

%\begin{table*}
\clearpage
\begin{landscape} 
%\begin{landscape}
\begin{table}
\begin{center}
\centering
\caption{Statistics of our blue straggler samples and comparisons to theoretical models.  The last four columns are taken directly from \citet{harris96}, and correspond to, respectively, the central velocity dispersion (in km s$^{-1}$), the absolute cluster $V$-band magnitude, the logarithm of the central cluster luminosity density (in units of L$_{\rm \odot}$ pc$^{-3}$) and the logarithm of the core relaxation time (in years).} 
\begin{tabular}{|c|c|c|c|c|c|c|c|c|c|c|c|c|c|c|c|}
\hline
%Cluster ID      &     $N_{\rm BS}$  &  \multicolumn{10}{|c|}{Comparisons to Numerical Scattering Experiments}     \\
%  (NGC)   &       &   \multicolumn{6}{|c|}{Velocity Distributions}  &  \multicolumn{4}{|c|}{Mass Distributions} \\
%         &        &    \multicolumn{2}{|c|}{1+2}   &  \multicolumn{2}{|c|}{2+2}   &  \multicolumn{2}{|c|}{Mass Transfer}   &  \multicolumn{2}{|c|}{1+2}  &  \multicolumn{2}{|c|}{2+2}  \\
%  &        &   KS statistic  &  P-value  &  KS statistic  &  P-value  &  KS statistic  &  P-value  &  KS statistic  &  P-value  &  KS statistic  &  P-value  \\
\hline
Cluster ID & $N_{\rm BS}$  & \multicolumn{4}{|c|}{Mass Distributions} &   \multicolumn{6}{|c|}{Velocity Distributions} & $\sigma_0$ (km s$^{-1}$) &  $M_V$  & log $\rho_0$ & log $\tau_{\rm rc}$ \\
(NGC)       &                        &  \multicolumn{2}{|c|}{1+2}  &  \multicolumn{2}{|c|}{2+2} &  \multicolumn{2}{|c|}{1+2}   &  \multicolumn{2}{|c|}{2+2}   &  \multicolumn{2}{|c|}{Mass Transfer}  &  &   &  &  \\
                 &                        &  KS statistic  &  P-value  &  KS statistic  &  P-value & KS statistic  &  P-value  &  KS statistic  &  P-value  &  KS statistic  &  P-value &   &  &  &  \\
\hline
3201     &  25  &  0.16 &   0.51  & 0.12 &  0.81   &  0.25 &  0.08    &   0.16  &  0.48   & 0.28 &  0.24   &  4.36  & -7.45 & 2.71   & 8.61 \\
4590     &  22  &  0.17 &   0.41  & 0.13 &  0.70   &  0.3  &  0.03    &   0.18  &  0.44   & 0.14 &  0.98   &  3.6   & -7.37 & 2.57   & 8.45  \\
4833     &  49  &  0.2  &   0.04  & 0.17 &  0.12   &  0.14 &  0.31    &   0.26  & 1.74e-3 & 0.29 &  0.03   &  6.05  & -8.17 & 3.00   &  8.78  \\
5286     &  22  &  0.3  &   0.03  & 0.3  &  0.48   &  0.42 & 3.72e-5  &   0.42  & 4.44e-4 & 0.32 &  0.17   &  9.07  & -8.74 & 4.10   &  8.40 \\
5927     &  31  &  0.16 &   0.36  & 0.19 &  0.21   &  0.09 &  0.96    &   0.13  &  0.66   & 0.45 & 2.21e-3 &  6.93  & -7.81 & 4.09   &   8.39 \\
6101     &  18  &  0.2  &   0.41  & 0.15 &  0.75   &  0.33 &  0.03    &   0.41  & 2.83e-3 & 0.44 &  0.04   &  2.77  & -6.94 & 1.65   &   9.21 \\
6121     &  16  &  0.22 &   0.38  & 0.28 &  0.14   &  0.73 & 1.82e-8  &   0.42  & 4.51e-3 & 0.31 &  0.35   &  5.14  & -7.19 & 3.64   &   7.90 \\
6171     &  28  &  0.35 & 1.55e-3 & 0.33 & 3.05e-3 &  0.21 &  0.13    &   0.18  &  0.29   & 0.14 &  0.92   &  4.28  & -7.12 & 3.08   &   8.06 \\
6218     &  52  &  0.29 & 2.88e-4 & 0.25 & 1.55e-3 &  0.09 &  0.76    &   0.17  &  0.08   & 0.19 &  0.26   &  5.22  & -7.31 & 3.23   &  8.19 \\
6254     &  45  &  0.33 & 5.65e-5 & 0.3  & 4.70e-4 &  0.18 &  0.09    &   0.31  & 2.03e-4 & 0.36 & 4.77e-3 &  5.37  & -7.48 & 3.54   &   8.21 \\
6304     &  29  &  0.09 &   0.95  & 0.1  &  0.9    &  0.11 &  0.87    &   0.21  &  0.15   & 0.21 &  0.51   &  6.15  & -7.30 & 4.49   &  7.36 \\
6341     &  31  &  0.12 &   0.72  & 0.12 &  0.7    &  0.25 &  0.03    &   0.31  & 2.89e-3 & 0.26 &  0.22   &  8.22  & -8.21 & 4.30   &  7.96 \\
6352     &  28  &  0.49 & 1.33e-6 & 0.46 & 8.14e-6 &  0.19 &  0.26    &   0.27  &  0.02   & 0.32 &  0.09   &  3.19  & -6.47 & 3.17   &  8.46 \\
6362     &  39  &  0.18 &   0.13  & 0.15 &  0.3    &  0.22 &  0.04    &   0.38  & 1.33e-5 & 0.18 &  0.51   &  2.89  & -6.95 & 2.29   &  8.80 \\
6366     &  8   &  0.37 &   0.18  & 0.36 &  0.21   &  0.75 & 9.99e-5  &   0.43  &  0.08   & 0.25 &  0.93   &  2.89  & -5.74 & 2.39   &   8.74 \\
6388     &  71  &  0.12 &   0.25  & 0.17 &  0.03   &  0.18 &  0.02    &   0.2   & 6.56e-3 & 0.37 & 9.36e-5 & 19.13  & -9.41 & 5.37   &   7.72 \\
6541     &  45  &  0.3  & 5.28e-4 & 0.25 & 1.55e-3 &  0.47 & 1.49e-9  &   0.07  &  0.96   & 0.36 & 4.77e-3 &  8.45  & -8.52 & 4.65   &   7.55 \\
6584     &  18  &  0.16 &   0.71  & 0.15 &  0.76   &  0.23 &  0.27    &   0.20  &  0.44   & 0.28 &  0.43   &  5.00  & -7.69 & 3.33   &  8.13 \\
6624     &  25  &  0.26 &   0.06  & 0.24 &  0.11   &  0.7  & 2.01e-11 &   0.14  &  0.65   & 0.28 &  0.24   &  7.19  & -7.49 & 5.30   &  6.61 \\
6637     &  32  &  0.22 &   0.07  & 0.22 &  0.08   &  0.27 &  0.02    &   0.13  &  0.62   & 0.25 &  0.24   &  7.31  & -7.64 & 3.84   &   8.15 \\
6652     &  15  &  0.29 &   0.12  & 0.27 &  0.19   &  0.23 &  0.34    &   0.19  &  0.61   & 0.2  &  0.89   &  0.19  & -6.66 & 4.48   &   7.05 \\
6656     &  32  &  0.21 &   0.11  & 0.18 &  0.24   &  0.14 &  0.51    &   0.19  &  0.19   & 0.19 &  0.59   &  8.27  & -8.50 & 3.63   &   8.53 \\
6681     &  13  &  0.49 & 2.40e-4 & 0.46 & 5.50e-3 &  0.31 &  0.13    &   0.15  &  0.9    & 0.31 &  0.49   &  5.2   & -7.12 & 5.82   &  5.82 \\
6717     &  15  &  0.23 &   0.34  & 0.19 &  0.38   &  0.4  &  0.01    &   0.51  & 3.48e-4 & 0.47 &  0.05   &  3.72  & -5.66 & 4.58   &   6.52 \\
6723     &  24  &  0.38 & 1.33e-3 & 0.37 & 1.87e-3 &  0.3  &  0.02    &   0.16  &  0.5    & 0.33 &  0.11   &  5.69  & -7.83 & 2.79   &   8.79 \\
6809     &  18  &  0.21 &   0.38  & 0.26 &  0.5    &  0.37 &  0.01    &   0.19  &  0.47   & 0.33 &  0.22   &  4.03  & -7.57 & 2.22   & 8.90 \\
6838     &  25  &  0.16 &   0.50  & 0.12 &  0.81   &  0.43 & 1.0e-4   &   0.22  &  0.16   & 0.28 &  0.24   &  3.27  & -5.61 & 2.83   &   7.54 \\
7089     &  22  &  0.28 &   0.05  & 0.29 &  0.07   &  0.24 &  0.15    &   0.17  &  0.52   & 0.14 &  0.98   &  9.04  & -9.03 & 4.00   &  8.48 \\
7099     &  26  &  0.17 &   0.41  & 0.13 &  0.7    &  0.57 & 3.3e-8   &   0.17  &  0.40   & 0.35 &  0.07   &  5.15  & -7.45 & 5.01   &  6.37 \\
\hline
\end{tabular}  
\end{center}
\label{table:stats}
\end{table}
\end{landscape}
\clearpage

Figures~\ref{fig:fig7}-~\ref{fig:fig10} show a comparison of the observed BS mass and velocity against the set of simulated distributions for NGC 4833, NGC 6362, NGC 6809 and NGC 6838, respectively. The sub-panels in each figure show a comparison of the cumulative distributions of the observed data against a Maxwellian velocity distribution (top), collision product velocity distribution (middle) and collision product stellar mass distribution (bottom). We obtain the $p$-value statistic from a 2-sided KS test in order to assess the agreement between the data and each theoretical prediction.

\begin{figure}
\begin{center}
\includegraphics[width=\columnwidth]{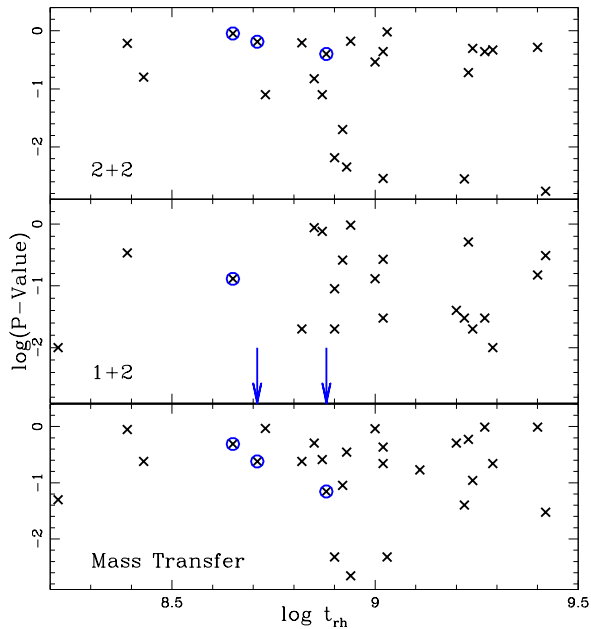}
\end{center}
\caption[$P$-values from our KS tests for each formation mechanism as a function of the logarithm of the cluster half-mass relaxation time]{The $p$-values obtained from our KS tests as a function of the logarithm of the cluster half-mass relaxation time \citep{harris96}, for every GC in Table~\ref{table:stats}.  We show the results for mass transfer, and collisions during 1+2 and 2+2 interactions in the bottom, middle and top panels, respectively. The open blue circles denote the post-core collapse (PCC) clusters in our sample.  The blue arrows indicate that the data points for two of the PCC clusters have $p$-values $<$ 10$^{-3}$.
\label{fig:fig12}}
\end{figure}

Figures~\ref{fig:fig11} and Figure~\ref{fig:fig12} show our results upon comparing all GCs in our samples.  Figure~\ref{fig:fig11} shows that in both panels for the 1+2 (bottom) and 2+2 (top) scenarios, a large number of points fall above the dashed line, indicating that MT is preferred over either collisional scenarios, while in many of these cases the corresponding 1+2 and/or 2+2 scenarios have very low p-values such that the null hypotheses for these models can be ruled out.  We also point out that more of the solid red points fall below the dashed line relative to the panel below showing the 1+2 $p$-values.  Hence, the 2+2 channel is preferred over the MT channel more often than is the case for the 1+2 scenario.  In particular, the velocity distributions for the 2+2 channel are shifted to higher velocities for most GCs, and more often they provide a better fit to the observed data than the comparatively slower 1+2 collision products.  Similarly, the calculated 2+2 mass distributions predict slightly higher masses relative to the 1+2 case, by a factor of $\sim 1.1$.  This is primarily due to the higher probability of collisions between more than two stars during four-body interactions relative to three-body interactions \citep[e.g.][]{leigh11,leigh12b}.

If most BSs are formed from collisions during 1+2 and 2+2 interactions, then the resulting BS velocity distribution will resemble the results of our Fewbody calculations immediately after formation.  Over time, however, relaxation will transform this initial distribution in to a Maxwellian.  Conversely, if BSs are formed from mass-transfer in binaries, then we expect the resulting velocity distribution to reflect that of the progenitor binaries, which has had sufficient time to become thermalized.  Hence, the binary mass transfer hypothesis for BS formation predicts a Maxwellian velocity distribution immediately after formation. Although the \textit{mean} BS lifetime is expected to exceed the relaxation time in the cluster core by up to 1-2 orders of magnitude for all but the most massive and extended GCs in our sample \citep{sills01,harris96}, most collisionally-produced BSs are more massive and have shorter MS lifetimes \citep{chatterjee13}.  Hence, while most BSs should have had sufficient time to relax dynamically, such that their velocity distribution resembles a Maxwellian independent of the preferred formation mechanism, this may not be the case for all BSs. Below we discuss in more detail one such scenario.

A few interesting scenarios could occur that are relevant to the collisional channel for BS formation.  For example, if a GC recently passed through a phase of deep core collapse, then the temporarily elevated central density would contribute to a sharp increase in the rates of collisions during 1+2 and 2+2 interactions.  For approximately a relaxation time after core collapse, these new collisionally-formed BSs should adhere to a velocity distribution decided by energy and momentum conservation during individual 1+2 and 2+2 interactions, as calculated by our numerical scattering simulations for every GC in our sample.  In such an extreme case, the observed BS population could indeed have an observed velocity distribution that deviates from a Maxwellian distribution, and resembles more closely the velocity distribution predicted by our numerical scattering simulations.  With that said, this particular example certainly requires some fine-tuning, and we would not expect to observe more than a few GCs in our sample in such a phase of their lifetime \citep[e.g.][]{fregeau09}.  To check this, we note that three GCs in our BS velocity samples are suspected to be in a post-core collapse phase, namely NGC 6624, NGC 6681 and NGC 7099.  In all three clusters, the $p$-values for the 2+2 collision scenario are the highest, and are even higher than the $p$-values obtained for the Maxwellian velocity distributions corresponding to the mass transfer scenario (see Table~\ref{table:stats}).  The $p$-values for the 1+2 and binary MT channels are low enough (i.e., $\lesssim 0.1$) for all but one PCC cluster, such that the null hypothesis that these models describe the data should be rejected.  More specifically, for NGC 6624 the 1+2 collision scenario can be rejected at high confidence. For NGC 6681, none of the scenarios can be rejected at high confidence, although the 1+2 collision scenario is the weakest at a $p$-value of $\sim 0.1$. The most notable case is NGC 7099, for which the 1+2 collision scenario can be rejected at high confidence and the mass transfer scenario can be rejected at modest confidence, hence making this GC the strongest candidate in our sample for having a large fraction of BSs formed via 2+2 collisions.

To further test these competing hypotheses, we search for possible correlations between the $p$-values for each formation channel provided in Table~\ref{table:stats} and the half-mass relaxation time, $t_{\rm rh}$.  Figure~\ref{fig:fig12} shows that there is no strong correlation between the $p$-values obtained for the mass transfer channel and the cluster half-mass relaxation time.  If such a correlation were present,\footnote{Or, more accurately, some systematic trend at low $p$-values satisfying $p \le 0.1$, since $p$-values are most reliable in this regime.} one interpretation would be that the \textit{initial} BS velocity distribution immediately after formation is \textit{not} Maxwellian, but instead evolves towards a dynamically-relaxed state over a relaxation time. This would, in principle, be more consistent with the collisional hypothesis, since the binary mass-transfer hypothesis predicts that BSs descended directly from internal evolution within (relatively) old binaries that have had plenty of time to relax toward a Maxwellian velocity distribution. Thus, the results from this component of our analysis are most consistent with a mass-transfer origin for the majority of the BSs in our samples, since the observed BS velocity distributions are typically consistent with having formed from a dynamically-relaxed population. However, as already discussed, this is insufficient to fully exclude the collisional channel, because collisions among the dynamically-relaxed binary population would yield the same Maxwellian velocity distribution of the collision products in very relaxed clusters. As shown in Table~\ref{table:stats} and Figure~\ref{fig:fig12}, the 2+2 collisional velocity distributions best match the observed BS velocity distributions for the post-core collapse GCs in our sample, namely NGC 6624, NGC 6681 and NGC 7099.  The latter being our strongest candidate for having a large fraction of BSs formed via 2+2 collisions. This provides good independent evidence to support the claim of a strong contribution from collisions to BS formation in NGC 7099, as found by \citet{ferraro09}. 
%As further supported by Figure~\ref{fig:fig11}, and Figures~\ref{fig:fig7}-\ref{fig:fig10}, the binary mass transfer hypothesis is preferred for the majority of clusters in our sample.  
%TERESA - CAN YOU PLEASE CHECK FOR A CORRELATION BETWEEN THE MT P-VALUES AND THE CLUSTER HALF-MASS AND CORE RELAXATION TIMES?  DEPENDING ON IF A CORRELATION IS PRESENT, WE CAN USE THIS TO HELP FURTHER CONSTRAIN THE PREFERRED FORMATION SCENARIO (SEE THE DISCUSSION SECTION).

\section{Summary and Discussion} \label{discussion}

In this paper, we have performed a thorough statistics-based search to answer the question:  What are the origins of BS stars in GCs?  The BS catalog of \citet{simunovic16} provides us with observational measurements for BS velocities (from their proper motions) and masses (from isochrone-fitting and assuming they are single isolated objects).  We compare these observed BS masses and velocities to analogous theoretical predictions for BSs formed from MS-MS collisions during single-binary (1+2) and binary-binary (2+2) interactions, obtained via a suite of numerical scattering experiments performed with the \texttt{FEWBODY} code \citep{fregeau04}.  

The results of this analysis are ultimately consistent with the BS samples having been derived from a dynamically-relaxed population.  At face value, this appears to favour the mass-transfer (and/or binary merger) hypothesis for BS formation. This is because the p-values obtained from Kolmogorov-Smirnov tests suggest that the observed BS velocity distributions most closely resemble Maxwellian velocity distributions, adjusted to account for the slightly higher masses of the progenitor binaries.  That is, we typically find lower p-values for the 1+2 and 2+2 channels relative to the MT channel, which suggests that for these cases (i.e., small p-values, or $p \lesssim 0.1$) we can reject the null hypothesis that the observed data are drawn from the models.  

Importantly, a Maxwellian velocity distribution is insufficient to generally rule out the collisional BS formation channel.  The reason is that the central relaxation timescales of our sample GCs are typically sufficiently short relative to the expected BS lifetime.  Therefore, even for collisional BSs, their observed velocities would be dynamically relaxed at the present-day cluster age.  More recently formed collisional BSs would have 2-D velocities that deviate from a Maxwellian velocity distribution, as shown from our numerical scattering experiments. In at least a handful of GCs in our sample, it is not unreasonable to expect that such a signature could be present due to a rapid burst of BS formation occurring due to collisions during an episode of core collapse.  In fact, for all post-core collapse GCs in our sample, we find that the observed BS velocities are best fit by the predicted 2+2 collision velocity distributions, favouring the collisional hypothesis for BS formation.
%Nevertheless, we do not find any strong evidence for such a signature in the observed 2-D BS velocity distributions, and ultimately that direct MS-MS collisions contribute significantly to BS formation in Galactic GCs.

No strong correlation exists between the p-values obtained for the mass transfer channel and the cluster half-mass relaxation time.\footnote{We emphasize that a \textit{correlation} is, in this case, not exactly the relevant flag to search for.  Instead, we look for something systematic in the distribution of low p-values with $p < 0.1$ (since larger p-values are less informative).  For example, if clusters with long half-mass relaxation times had only low p-values, which is not observed in our data.}  If such a correlation were present, one interpretation would be that the \textit{initial} BS velocity distribution immediately after formation is \textit{not} Maxwellian, but instead evolves toward this steady-state significantly over a relaxation time.  This would, in principle, be more consistent with the collisional hypothesis, since the binary mass-transfer and merger hypothesis predicts that BSs are descended directly from internal evolution within (relatively) old binaries that have had plenty of time to relax toward a Maxwellian velocity distribution.  Thus, the results from this component of our analysis primarily identify a lack of evidence for the collisional channel for the majority of the BSs in our samples, with the possible exception of post-core collapse clusters, since the observed BS velocity distributions are consistent with having formed from a dynamically-relaxed population.  However, for the three GCs in our BS velocity samples suspected of being in a post-core collapse phase, namely NGC 6624, NGC 6681 and NGC 7099, the preferred agreement with the 2+2 collisional velocity distributions could be interpreted as evidence that the rate of BS formation due to collisions is enhanced during core-collapse.  We note that previous studies have speculated about the connection between core-collapse and BS formation by collisions. In particular, collision-product isochrones have been used to suggest the collisional origin of sub-populations of BSs in NGC 7099 \citep{ferraro09} (which is also our strongest candidate for collisions being the dominant mechanism for BS formation), NGC 362 \citep{dalessandro13} and NGC 1261 \citep{simunovic14} (unfortunately, neither of the last two GCs are in our sample). Our results therefore seem to be consistent with this hypothesis.

We do not use the distribution of inferred BS masses from our analysis, since we do not know the underlying binarity of each BS.  Consequently, we defer this exercise to future work.  If a more detailed and thorough SED-analysis is performed that can reveal the single or binary nature of each BS in our sample, this would allow us to return to this component of our analysis and re-perform it with significant refinements informed by our initial analysis. We hope to perform this SED-fitting exercise to constrain the binarity of each BS in a forthcoming paper.

Our primary conclusions can be summarized as follows:

\begin{itemize}

\item The results of our BS velocity distribution analysis suggest that each of the 1+2, 2+2 and binary MT channels are preferred by at least some clusters.  The majority of the clusters in our sample prefer the MT hypothesis over either collisional hypotheses, as shown in Figures~\ref{fig:fig11} and~\ref{fig:fig12}.

\item Of those clusters that prefer a collisional origin during our BS velocity distribution analysis, most prefer the 2+2 collisional velocity distributions over the 1+2 distributions.

\item Finally, we note that all three of the post-core collapse clusters included in the BS velocity distribution analysis (i.e., NGC 6624, NGC 6681 and NGC 7099) prefer the 2+2 collisional hypothesis over either the 1+2 or MT hypotheses.

%\item \textbf{The methods presented in this paper for comparing observed BS SEDs to theoretical templates can be used in future studies.  However, the data used in this paper to performed the SED analysis should be regarded with caution.  Specifically, the flux uncertainties used in this analysis were calculated somewhat inhomogeneously, and there is no reason to expect that they are appropriate for the analyses performed in this paper.  Indeed, our results are consistent with these uncertainties being under-estimates of their true values (due to effects such as variability from pulsations, eclipses in close binaries, etc.), as shown via the addition of an intrinsic dispersion term in our calculation of the reduced chi-squared values.}
%
%\item \textbf{After including an intrinsic dispersion term in the calculations of the observed flux uncertainties there are several clusters for which we find a larger number of RGB and MSTO objects well-fit by a single star SED relative to the corresponding BS samples, as shown in the last three columns of Table~\ref{table:stats3}. These GCs are NGC 1261, NGC 5927, NGC 6101, NGC 6388 and NGC 6681.}

\end{itemize}

In a forthcoming paper, we will also use a similar procedure as outlined in this paper to check for low-mass collision products below the main-sequence turn-off, as illustrated in Figure~\ref{fig:fig1}.  Both collisionally-formed BSs during 1+2 and 2+2 interactions and MS-MS binary mergers are expected to resemble the evolutionary tracks provided in Figure~\ref{fig:fig1} immediately after the collision/merger.  Several Gyrs later, these "dormant'' collision/merger products hiding on the MS could appear brighter and bluer than the main-sequence turn-off, coinciding with BSs \citep[e.g.][]{hypki13}.  By constraining the numbers of such low-mass collision/merger products in the CMD and convolving these numbers with stellar evolution models (which quantify the rate at which objects should evolve in colour and brightness, and hence move around in the CMD), they can be used to constrain the rates of MS-MS mergers and/or collisions.  In turn, these constraints can be extrapolated to other regions of the CMD, especially BSs, in order to constrain the expected contribution from collisions/mergers.

\section*{Acknowledgments}

N.W.C.L. acknowledges support from a Kalbfleisch Fellowship at the American Museum of Natural History.  T.H.P. acknowledges the support through the FONDECYT Regular Project No. 1161817 and the BASAL Center for Astrophysics and Associated Technologies (PFB-06).

\end{document}